\documentclass{elsart}
\usepackage{natbib}
\usepackage{amsfonts}
\usepackage{amssymb}
\usepackage{psfig}

\begin{document}

\newlength{\figwidtha}
\newlength{\figwidthb}
\setlength{\figwidtha}{0.5\linewidth}
\setlength{\figwidthb}{\linewidth}
\setlength{\textwidth}{150mm}
\setlength{\textheight}{240mm}
\setlength{\parskip}{2mm}

\input{epsf.tex}
\epsfverbosetrue

\renewcommand{\baselinestretch}{1.0}
\def\ao{Appl.\  Opt.\ }
\def\ap{Appl.\  Phys.\ }
\def\apl{Appl.\ Phys.\ Lett.\ }
\def\apj{Astrophys.\ J.\ }
\def\bell{Bell Syst.\ Tech.\ J.\ }
\def\jqe{IEEE J.\ Quantum Electron.\ }
\def\assp{IEEE Trans.\ Acoust.\ Speech Signal Process.\ }
\def\aprop{IEEE Trans.\ Antennas Propag.\ }
\def\mtt{IEEE Trans.\ Microwave Theory Tech.\ }
\def\iovs{Invest.\ Ophthalmol.\ Vis.\ Sci.\ }
\def\jcp{J.\ Chem.\ Phys.\ }
\def\jmo{J.\ Mod.\ Opt.\ }
\def\josa{J.\ Opt.\ Soc.\ Am.\ }
\def\josaa{J.\ Opt.\ Soc.\ Am.\ A }
\def\josab{J.\ Opt.\ Soc.\ Am.\ B }
\def\jpp{J.\ Phys.\ (Paris) }
\def\nat{Nature (London) }
\def\oc{Opt.\ Commun.\ }
\def\ol{Opt.\ Lett.\ }
\def\pl{Phys.\ Lett.\ }
\def\pra{Phys.\ Rev.\ A }
\def\prb{Phys.\ Rev.\ B }
\def\prc{Phys.\ Rev.\ C }
\def\prd{Phys.\ Rev.\ D }
\def\pre{Phys.\ Rev.\ E }
\def\prl{Phys.\ Rev.\ Lett.\ }
\def\rmp{Rev.\ Mod.\ Phys.\ }
\def\pspie{Proc.\ Soc.\ Photo-Opt.\ Instrum.\ Eng.\ }
\def\sjqe{Sov.\ J.\ Quantum Electron.\ }
\def\vr{Vision Res.\ }

\begin{frontmatter}
\title{Engineering of spatial solitons in two-period QPM structures} 
\author{Steffen Kj\ae r Johansen$^{*}$, Silvia Carrasco, and Lluis Torner}
\address{Laboratory of Photonics, Department of Signal Theory and
Communications,\\ Universitat Politecnica de Catalunya, \\ Gran Capitan
UPC-D3, Barcelona, ES 08034, Spain}
\author{Ole Bang}
\address{Department of Informatics and Mathematical Modelling, \\ Technical University 
of Denmark, DK-2800 Lyngby, Denmark}

\maketitle

\normalsize

\begin{abstract}
We report on a scheme which might make it \emph{practically} possible to
engineer the effective competing nonlinearities that on average govern the
light propagation in quasi-phase-matching (QPM) gratings. Modulation of the QPM period with a second longer period,
introduces an extra degree of freedom, which can be used to engineer
the effective quadratic and  induced cubic nonlinearity. However, in
contrast to former work here we use a \emph{simple phase-reversal
grating} for the modulation, which is practically realizable and has
already been fabricated. Furthermore, we develop the theory for \emph{arbitrary relative lengths 
of the two periods} and we consider the effect on solitons and the 
bandwidth for their generation.
We derive an expression for the bandwidth of multicolor soliton generation 
in two-period QPM samples and we predict and confirm numerically that the 
bandwidth is broader in the two-period QPM sample than in 
homogeneous structures.
\end{abstract}

\end{frontmatter}

\normalsize

Quasi-phase-matching (QPM) is a major alternative over conventional phase
matching in many laser applications based on frequency-conversion processes in
quadratic nonlinear media (for reviews, see \cite{bye97,fej98beamshaping}).
Besides other practical advantages, QPM allows tailoring the nonlinearity of
the material to form complex structures. This opens a range of
new possibilities, which have become experimentally feasible with the recent
progress in poling techniques.  For example, engineerable pulse
compression in frequency-doubling schemes in synthetic QPM gratings has been
demonstrated in aperiodically poled lithium niobate and potassium niobate
\cite{arbore97janol,arbore97sepol,loza-alvarez99ol,imeshev00febjosab}, and 
transverse QPM gratings have been
made both for shaping second-harmonic beams and to extend the spectral coverage
of optical parametric oscillators\cite{imeshev98mayol,powers98ol}. Bandwidth
enhanced parametric interactions can be obtained in modulated-period structures
\cite{mizuuchi94ieee}, multiple nonlinear interactions can be achieved in
quasi-periodic schemes \cite{zhu97octscience,keren99nov99,chou99augol,liu01augapl}, and simultaneous
generation of multiple color laser light has been demonstrated in QPM crystals
doped with  active lasing ions \cite{capmany01janapl}. QPM engineering
also finds novel important applications beyond pure frequency-conversion devices,
e.g., to generate enhanced cascading phase-shifts \cite{cha98febol},
all-optical diode operation \cite{gallo01julapl}, and multicolor soliton
formation
\cite{torner99decopn,torner98junol,clausen97junprl,clausen99janol,clausen99decprl,carrasco00sepol}.

Here we show that two-period QPM structures offer new opportunities for 
soliton control.

Multicolor solitons mediated by second-harmonic generation (SHG) form by mutual
trapping of the beams at the fundamental frequency (FF) and at the second
harmonic frequency (SH).  Here we focus on bright solitons whose basic
properties are well known. Whole dynamically stable families exist above a
certain power threshold for all phase mismatches.  At lowest order the effect
of QPM grating is to average the quadratic nonlinearity
\cite{fej98beamshaping}. However, taking higher order perturbations into account reveals that the 
corresponding averaged field equations include effective cubic, Kerr-like 
terms\cite{clausen97junprl,bang99ol,corney01seppre}. Such terms modify
the average properties of CW waves \cite{kobyakov98aprol,bang01julol} and the soliton families of the averaged 
equations \cite{clausen97junprl,corney01seppre,corney01sepprl},
which can be analyzed as sustained by competing quadratic and 
effective cubic nonlinearities \cite{buryak95octol,bang97janjosab,berge97marpre,bang98oc,bang98pre}.

Nevertheless, in standard QPM the strength of the induced averaged cubic
nonlinearity is proportional to the ratio between the QPM grating period and
the soliton characteristic length (i.e., the diffraction length in the case of
spatial solitons). At optical wave lengths, the former is typically of the
order of ten $\mu$m, whereas the latter is a few mm. 
Therefore, the strength of the induced cubic nonlinearities is
extremely small. 
The question thus naturally arises whether and how QPM engineering can be 
employed to bring the effective cubic nonlinearities to compete with the 
quadratic in the average fields. One solution is to add a strong
dc-part to the nonlinear QPM grating, i.e. as done in
\cite{helmy00sepol,shuto97marieee}. This adds a term to the induced
Kerr terms, which is proportional to the QPM grating period and the
dc-value squared, and thus can be large\cite{corney01seppre}. Another
potentially more versatile technique is to modulate the QPM period
with a second longer period, as it was shown theoretically in
\cite{bang99ol}. This introduces an extra degree of freedom, which can
be used to engineer the effective quadratic and induced averaged cubic
nonlinearity.

Here we consider the latter technique. However, in contrast to the rather 
complicated long-period modulation used in \cite{bang99ol} we use here a 
\emph{simple phase-reversal grating} for the modulation, similar to 
the two-period QPM sample recently employed by Chou and co-workers 
\cite{chou99augol} for multiple-channel wavelength conversion in the 
third telecommunication window.
Furthermore, we develop the theory for \emph{arbitrary relative lengths 
of the two periods} and we consider the effect of the modulation on 
multicolor solitons.
We expect soliton formation to be possible for a variety of phase-reversal
periods, in view of the fact that multicolor solitons have been shown to
be robust against strong  
perturbations, including  quasi-periodic QPM
gratings\cite{clausen99decprl} and even periodic gain and
loss\cite{torner00novoc}. Using the induced averaged cubic nonlinearities we derive an expression for the bandwidth of multicolor soliton generation in two-period QPM samples and 
we show that the bandwidth is broader in the two-period QPM 
sample than in homogeneous quadratic nonlinear materials.
Importantly, all our results are confirmed numerically.

We consider beam propagation under type-I SHG conditions in a lossless
QPM $\chi^{(2)}$ slab waveguide. The slowly varying envelope of the
fundamental wave, $E_1=E_1(x,z)$, and its SH, $E_2=E_2(x,z)$,
are\cite{menyuk94}
\begin{eqnarray}
&&i\frac{\partial E_1}{\partial z}+\frac{1}{2}\frac{\partial^2 E_1}{\partial x^2}+d(z)E_1^{*}E_2e^{-i\beta z}=0,\label{eqn:chi23system1}\\
&&i\frac{\partial E_2}{\partial z}+\frac{\alpha}{2}\frac{\partial^2 E_2}{\partial x^2}+d(z)E_1^{2}e^{i\beta z}=0,\label{eqn:chi23system2}
\end{eqnarray}
where in all cases of practical interest $\alpha\simeq 0.5$.
 The normalized wave-vector mismatch is introduced via the real parameter
$\beta=k_1\omega_0^2\Delta k$, where $\Delta k=2k_1-k_2$, and $\omega_0$ is the beam
width, and $k_{1,2}$ are the linear wave numbers of the fundamental and SH,
respectively. The scaled transverse coordinate, $x$, is measured in units of
$\omega_0$ and the propagation coordinate, $z$, is measured in units of $2l_d$
where $l_d=k_1\omega_0^2/2$ is the diffraction length of the fundamental wave. The
spatial periodic modulation of the nonlinearity is described by the grating
function $d(z)$ whose amplitude is normalized to 1, and whose domain length we
define as $\Lambda=\pi/\kappa$, where $\kappa$ is the spatial grating
frequency or the QPM frequency, which we assume real and positive. In the case of second-harmonic generation in lithium niobate pumped
at $\lambda_{\omega}\sim$ 1.5 $\mu$m with a beam width of $\omega_0\sim 20$ $\mu$m,
the intrinsic material wave vector  mismatch is of the order $|\beta|\sim
10^3$.  Thus the QPM frequency must also be of this order which
corresponds to a domain length of $\sim10$ $\mu$m. For such values, a scaled
grating frequency of the order $\kappa\sim10$ correspond to a domain length of
about 1 mm.

\begin{figure}
\centerline{\hbox{
\psfig{figure=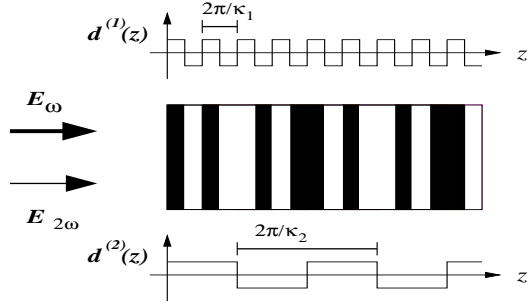,width=7cm,height=4cm}
}}\vspace{0.5cm}
\caption{Two-period QPM grating.}
\label{fig:crystal}
\end{figure}

\begin{figure}
\centerline{\hbox{
\psfig{figure=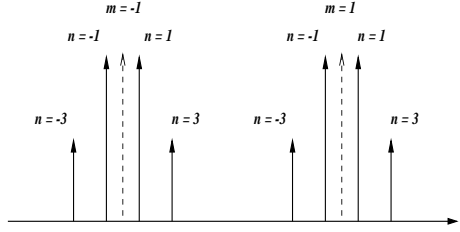,width=6cm,height=3cm}
}}\vspace{0.5cm}
\caption{Peak splitting in the two-period QPM grating. Dashed peaks
indicate the location of the 1'st order peaks in the one-period case.}
\label{fig:peaks}
\end{figure}

The two-period grating function, $d(z)$, consists of a primary grating,
$d^{(1)}(z)$, and a superimposed secondary grating, $d^{(2)}(z)$. We
expand $d(z)$ in a Fourier series
\begin{eqnarray}\label{eqn:gennonlin}
d(z)=\sum_k d_k^{(1)}\exp(ik\kappa_1 z)\times\sum_l d^{(2)}_l\exp(il\kappa_2 z)
\end{eqnarray}
where the summations are over all $(k,l)$ from $-\infty$ to $\infty$. If we
assume the grating functions to be square, only the odd harmonics
enters into the expansion, $d_{2l+1}=2/i\pi(2l+1)$ and
$d_{2l}=0$. In the case of low-depletion SHG, the effect of the superimposed
period is to split each peak of the original one-period QPM grating into an
infinite family of peaks\cite{chou99augol}. More formally, one has peaks at all spatial QPM frequencies
$m\kappa_1+n\kappa_2$ where $m$ and $n$ are the QPM orders related to the
primary and secondary grating, respectively. These peaks are not
delta-like, though for simplicity they are depicted as such in
Fig. \ref{fig:peaks}, but rather sinc-like functions around the relevant QPM
frequency. The SHG-efficiency is higher the
lower the order. Hence, in  one-period QPM  one would choose to 
match to the first peak $(m=\pm1)$. In two-period QPM there is an
equivalent rule, i.e. the low-depletion SHG-efficiency is highest for $m=n=\pm
1$. To be able to treat the problem mathematically we have to neglect
all overlap between peaks in order to avoid resonances. Thus we must
assume $\kappa>>1$ in the one-period case.  In the general
two-period case the spectrum becomes
dense, as in the case of a Fibonacci grating\cite{clausen99decprl}. In
order to be able to neglect any overlap between the peak we are
looking at, and the rest of the dense spectrum, we have to assume, not
only high spatial QPM frequencies, but also that the two frequencies
are of different order. We remark that soliton excitation is expected to be
possible no matter what the spectrum looks like, as long as the power
is high enough\cite{clausen99decprl}. Our numerical simulations
presented here agree with this expectation. It is the lack of analytical
tools which necessitates the assumption of well separated peaks.

By applying the asymptotic expansion\cite{kivshar94pre} technique we have established a
perturbation theory describing the propagation of the averaged fields
in this two-period QPM system. We make the transformation
$E_1(x,z)=w(x,z)$ and $E_2(x,z)=v(x,z)\exp{(i\epsilon z)}$, where $\epsilon=\beta-m\kappa_1-n\kappa_2$, is the residual phase
mismatch, which is assumed to be small. The functions $w(x,z)$ and
$v(x,z)$ are assumed to vary slowly on the
scale given by the QPM periods and can be expanded in the Fourier series

\begin{eqnarray}
&&w(x,z)=w_0(x,z)+\sum_{k\neq0}w_k(x,z)e^{ik\kappa_1
z}+\sum_{l\neq0}\omega_l(x,z)e^{il\kappa_2
z},\label{eqn:newexpansion1}\\
&&v(x,z)=v_0(x,z)+\sum_{k\neq0}v_k(x,z)e^{ik\kappa_1
z}+\sum_{l\neq0}\nu_l(x,z)e^{il\kappa_2 z}.\label{eqn:newexpansion2}
\end{eqnarray}
We remark that this assumption is valid also when the QPM frequencies
share common harmonics. In this case
the SHG spectrum discussed above becomes discrete as in the one-period
case. However, to avoid overlap between peaks the QPM frequencies must
still be assumed of different order.  Furthermore we formally assume that $\beta>>1$, $\kappa_1>>1$, and $\kappa_2>>1$. Substituting
(\ref{eqn:newexpansion1}-\ref{eqn:newexpansion2}) into
(\ref{eqn:chi23system1}-\ref{eqn:chi23system2}) and matching leading
order terms yields the relations  for the
harmonic coefficients\cite{clausen97junprl}. One gets
\begin{eqnarray}
&&w_q=\frac{1}{q\kappa_1}   d_n^{(2)}d^{(1)}_{(q+m)}w_0^*v_0,\;\; 
v_n=\frac{1}{q\kappa_1}d_{-n}^{(2)}d^{(1)}_{(q-m)}w_0^2,\label{sys:harmonics1}\\
&&\omega_q=\frac{1}{q\kappa_2}   d_m^{(1)}d^{(2)}_{(q+n)}w_0^*v_0,\quad
\nu_n=\frac{1}{q\kappa_2}d_{-m}^{(1)}d^{(2)}_{(q-n)}w_0^2,\label{sys:harmonics2}
\end{eqnarray}
and the equations governing the averaged fields are of the form
\begin{eqnarray}
&& i\frac{\partial w_0}{\partial z}+\frac{1}{2}\frac{\partial^2
  w_0}{\partial x^2}
  +\eta w_0^{*}v_0
  +\gamma(|w_0|^2 - |v_0|^2)w_0=0,\label{eqn:avgQPMsystem1}\\
&& i\frac{\partial v_0}{\partial z}+\frac{1}{4}\frac{\partial^2
  v_0}{\partial x^2}
  -\epsilon v_0
  +\eta w_0^{2}
  -2\gamma|w_0|^{2}v_0=0.\qquad\label{eqn:avgQPMsystem2}
\end{eqnarray}
System (\ref{eqn:avgQPMsystem1}-\ref{eqn:avgQPMsystem2}) conserves the same
power as system (\ref{eqn:chi23system1}-\ref{eqn:chi23system2}),
$P=\int(|w_0|^2+|v_0|^2)dx=\int(|E_1|^2+|E_2|^2)dx$. The strengths of the
averaged nonlinearities are given by
\begin{eqnarray}
&&\eta=
d_m^{(1)}d_n^{(2)}-\sum_{k}d^{(1)}_{k}d^{(2)}_{[(k-m)\frac{\kappa_1}{\kappa_2}+n]},\label{eqn:eta}\\
&&\gamma=\frac{d_{-n}^{(2)}}{\kappa_1}\sum_{q,l}
\frac{d_l^{(2)}d_{[(m+q)-(l-n)\frac{\kappa_2}{\kappa_1}]}^{(1)}d^{(1)}_{[m+q]}}{q}+\frac{d_{-m}^{(1)}}{\kappa_2}\sum_{q,k}
\frac{d_k^{(1)}d_{[(n+q)-(k-m)\frac{\kappa_1}{\kappa_2}]}^{(2)}d^{(2)}_{[n+q]}}{q}.\hspace{1cm}\label{eqn:gamma}
\end{eqnarray}
In these expressions, $q$, $l$, and $k$ are integers, and the
summations are everywhere over all integers from $-\infty$ to $\infty$
except zero. We emphasize that the averaged model
(\ref{eqn:avgQPMsystem1}-\ref{eqn:avgQPMsystem2}) is valid for arbitrary values
of $\kappa_1$ and $\kappa_2$ as long as the assumptions
$\kappa_1>>1$, and $\kappa_2>>1$ are not violated. However care must be taken when calculating
the sums in (\ref{eqn:eta}) and (\ref{eqn:gamma}). When necessary, 
$k$ and $l$, themselves integers, must always be
chosen such that the $(k-m)\kappa_1/\kappa_2$ and $(l-n)\kappa_2/\kappa_1$ are
integers. This ensures that the Fourier coefficients  $d_k^{(1)}$ and
$d_l^{(1)}$ exist. We remark that if one of the QPM frequencies is an
irrational number expressions (\ref{eqn:eta}) and (\ref{eqn:gamma})
are greatly simplified
\begin{eqnarray}
&&\eta=
d_m^{(1)}d_n^{(2)},\label{eqn:eta1}\\
&&\gamma=\frac{d_{n}^{(2)}d_{-n}^{(2)}}{\kappa_1}\sum_{q}
\frac{d_{[m+q]}^{(1)^{\scriptstyle 2}}}{q}+\frac{d_{m}^{(1)}d_{-m}^{(1)}}{\kappa_2}\sum_{q}\frac{{d_{[n+q]}^{(2)^{\scriptstyle 2}}}}{q}.\label{eqn:gamma1}
\end{eqnarray}
To make the cubic self- and cross-phase modulation symmetric in
(\ref{eqn:avgQPMsystem1}-\ref{eqn:avgQPMsystem2}) we have had to assume that
$d^{*(i)}_{n}=d^{(i)}_{-n}=-d^{(i)}_{n}$, which is the case for square grating
functions. Otherwise expressions (\ref{eqn:eta}-\ref{eqn:gamma1}) hold for any
grating function of the form (\ref{eqn:gennonlin}). Rewriting
(\ref{eqn:eta}) and (\ref{eqn:gamma}) for square grating functions we get
\begin{eqnarray}
&&\eta=
-\frac{4}{\pi^2}\left(\frac{1}{mn}-\sum_{s}\frac{1}{(2s+m)\left(2s\frac{\kappa_1}{\kappa_2}+n\right)}\right),\label{eqn:squarenl1}\\
&&\gamma=-\frac{16}{\pi^4}\left(\frac{1}{n\kappa_1}\sum_{q,r}
\frac{1}{2q(2r+n)\left(2q-2r\frac{\kappa_2}{\kappa_1}+m\right)(m+2q)}\right.\nonumber\\
&&\hspace{2cm}\left.+\frac{1}{m\kappa_2}\sum_{q,s}
\frac{1}{2q(2s+m)\left(2q-2s\frac{\kappa_1}{\kappa_2}+n\right)(n+2q)}\right),\label{eqn:squarenl3}
\end{eqnarray}
where the summations now are over all integers $q$, $r$, and $s$ from
$-\infty$ to $\infty$ except zero. Notice the different sign of the
lowest-order contribution to (\ref{eqn:squarenl1}) at the ($m=1, n=1$)-peak,
and the ($m=1, n=-1$)-peak. Such change of sign has implications to the soliton
features, as shall be discussed below. Equations
(\ref{eqn:avgQPMsystem1}-\ref{eqn:avgQPMsystem2}) were first derived in \cite{clausen97junprl}. They have the same form regardless of the specific type of grating, the 
parameters still merely being given as sums over the Fourier coefficients of 
the grating. Thus they were derived in \cite{corney01seppre} for QPM
with both a linear and a nonlinear grating with or without a dc-value
of the nonlinear grating. They were also derived in \cite{bang99ol} for a more
exotic two-period grating.

By letting $w_0(x,z)=\bar{w}(x,z)/\eta$ and
$v_0(x,z)=\bar{v}(x,z)/\eta$ we can reduce system
(\ref{eqn:avgQPMsystem1}-\ref{eqn:avgQPMsystem2}) to
\begin{eqnarray}
&& i\frac{\partial \bar{w}}{\partial z}+\frac{1}{2}\frac{\partial^2
  \bar{w}}{\partial x^2}
  +\bar{w}^{*}\bar{v}
  +\tilde{\gamma}(|\bar{w}|^2 - |\bar{v}|^2)\bar{w}=0,\label{eqn:redavgQPMsystem1}\\
&& i\frac{\partial \bar{v}}{\partial z}+\frac{1}{4}\frac{\partial^2
  \bar{v}}{\partial x^2}
  -\epsilon \bar{v}
  +\bar{w}^{2}
  -2\tilde{\gamma}|\bar{w}|^{2}\bar{v}=0.\qquad\label{eqn:redavgQPMsystem2}
\end{eqnarray}
The stationary bright solitary families of system
(\ref{eqn:redavgQPMsystem1}-\ref{eqn:redavgQPMsystem2}) are of the form
$\bar{w}(x,z)=u_1(x)\exp(i\lambda  z)$, $\bar{v}(x,z)=u_2(x)\exp(i2\lambda z)$.
They are para\-metrized by the soliton parameter
$\lambda>\max\{0,-\epsilon/2\}$.
The stationary solutions of (\ref{eqn:avgQPMsystem1}-\ref{eqn:avgQPMsystem2})
are para\-metrized by exactly the same parameter but since this system is
characterized by three ($\epsilon$, $\eta$, and $\gamma$) rather than the two
parameters of (\ref{eqn:redavgQPMsystem1}-\ref{eqn:redavgQPMsystem2})
($\epsilon$ and $\tilde{\gamma}=\gamma/\eta^2$) each solution of
(\ref{eqn:redavgQPMsystem1}-\ref{eqn:redavgQPMsystem2}) corresponds to an
infinite family of solutions of
(\ref{eqn:avgQPMsystem1}-\ref{eqn:avgQPMsystem2}), i.e. an infinite number of
QPM frequency combinations and thus physical setups. System
(\ref{eqn:redavgQPMsystem1}-\ref{eqn:redavgQPMsystem2}) conserves the
renormalized power
$\bar{P}=\int(|\bar{w}|^2+|\bar{v}|^2)dx=\eta^2P$. We note that also the
assumed slow variation of the functions $w(x,z)$ and $v(x,z)$ imposes
constraints on the choice of the soliton parameter $\lambda$. The
characteristic soliton scale length, $2\pi/\lambda$, must be much larger than
the QPM period  
$\Lambda=\pi/\kappa$, where $\kappa$ in the
two-period case is the smallest QPM frequency. 

Merely looking at expressions (\ref{eqn:squarenl1}-\ref{eqn:squarenl3})
suggests that in principle there is ample room for engineering of the averaged
nonlinearities. The main parameters are the QPM frequencies $\kappa_1$ and
$\kappa_2$. In general we can state that the higher the QPM
frequencies are, the lower the induced averaged cubic terms become. Thus $\gamma\rightarrow0$ for
$\kappa_1,\kappa_2\rightarrow\infty$.  One also notices that the ratio between
$\kappa_1$ and $\kappa_2$ and vice versa plays an important role in the sums in
expressions (\ref{eqn:squarenl1}) and (\ref{eqn:squarenl3}). We observe
that the smaller the ratio the more significant the corresponding sums. From
(\ref{eqn:squarenl1}) it is obvious that the induced part of the averaged
quadratic nonlinearity depends
\emph{only} on the ratio $\kappa_2/\kappa_1$. Thus we can realize the same
averaged quadratic nonlinearity at large QPM frequencies (no induced
averaged cubic terms) and at low QPM frequencies (significant induced
averaged cubic terms). We note that if we
let the second grating function $d^{(2)}(z)=1$, the only Fourier coefficient left in the expansion of
$d^{(2)}(z)$ is $d^{(2)}_{0}=d^{(2)}_{n}=d^{(2)}_{-n}=1$ and hence
$\eta=d_m^{(1)}$ and $\gamma=\frac{1}{\kappa_1}\sum_{}
d_{[m+q]}^{(1)^{\scriptstyle 2}}/q$. The whole system thus degenerates
into the one-period case and the results of \cite{clausen97junprl} are reproduced.

\begin{figure}
\centerline{\hbox{
\psfig{figure=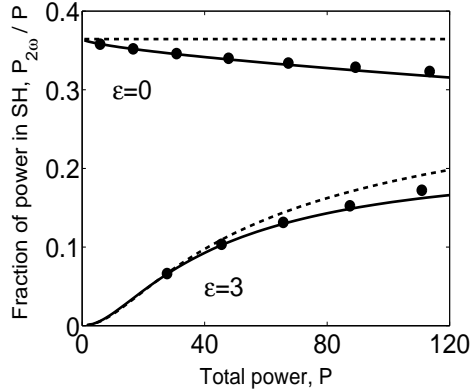,width=6cm,height=5cm}
}}\vspace{0.5cm}
\caption{Fraction of power in the SH for the two-period QPM system with
$(\kappa_1,\kappa_2)=(997,13)$ as a function of total power. Solid
line and $\bullet$: Theoretical and numerically measured values,
respectively. Dashed curves: Zero-order approximation, $\gamma=0$. The
value of $\epsilon$ is indicated at each pair of curves.}
\label{fig:i2w}
\end{figure}

Depending on our objectives we can make the individual terms of the averaged nonlinearities more or less significant. Thus we can essentially make
the two-period QPM system behave like the one-period system with induced
averaged cubic nonlinearities by choosing $\kappa_1>>\kappa_2$ while still
assuming $\kappa_2>>1$. In this way one can
use $\kappa_1$ to reduce a large intrinsic phase mismatch, $\beta$, to a small
effective mismatch which can then be manipulated further with
$\kappa_2$. Thus, if we let $\kappa_1\rightarrow\infty$ the averaged
nonlinearities of for the system with square grating functions
(\ref{eqn:squarenl1}-\ref{eqn:squarenl3}) with $m=n=1$ reduces to
\begin{eqnarray}
&&\eta=-\frac{4}{\pi^2},\qquad
\gamma=\frac{4}{\kappa_2\pi^2}\left(1-\frac{8}{\pi^2}\right)  
\end{eqnarray}
To verify the model in this $\kappa_1\rightarrow\infty$ limit we have made a
series of simulations launching soliton initial conditions, found from
(\ref{eqn:redavgQPMsystem1}-\ref{eqn:redavgQPMsystem2}), in a real
two-period structure described by
(\ref{eqn:chi23system1}-\ref{eqn:chi23system2}) with
$\kappa_1$ chosen to be a high prime and $\kappa_2$ a small prime. Choosing
primes effectively eliminates contributions from the remaining sums in
(\ref{eqn:squarenl1}-\ref{eqn:squarenl3}) because $r$ and $s$ must be
chosen such that $r\kappa_2/\kappa_1$ and 
$s\kappa_1/\kappa_2$ are positive integers. We measure the fraction of power in
the SH after any initial transient has died out and compare this to the
theoretically predicted value. In Fig. \ref{fig:i2w} we  have summarized the
results of this investigation for different powers and different effective
phase mismatches.

If both QPM frequencies are of the same order, the full
non-simplified expressions (\ref{eqn:squarenl1}-\ref{eqn:squarenl3}) must be
used to calculate the averaged nonlinearities. Because of the way the ratio
$\kappa_1/\kappa_2$ enters in (\ref{eqn:squarenl1}-\ref{eqn:squarenl3}), one
has \emph{discrete resonances} between the QPM frequencies. At these
discrete resonances the induced averaged nonlinearities become stronger. The
strength depends on the order of the resonance, the lower the order the higher
the strength. To show how the strength of the averaged nonlinearities are
calculated we concentrate on the summations involving
$s\kappa_1/\kappa_2$. The
contribution of the terms involving $r\kappa_2/\kappa_1$ is handled in an
analogous way. First-order resonances arise when $\kappa_1/\kappa_2$ is a
positive integer;  then $s$ is summed over all integers except zero from
$-\infty$ to $\infty$. Second-order resonances arise when
$\kappa_1=(2p-1)\kappa_2/2$ where $p$ is a positive integer. In this case $s$
only spans the even integers. Third-order resonances occur when
$\kappa_1=p*\kappa_2/3$ where $p=1,2,4,5,7,8,\cdots$; then $s$ must be summed
over the integers $\pm3,\pm6,\pm9,\cdots$. And so on and so forth for
higher-order resonances.

We emphasize that resonances in principle must be taken into account
whenever the gratings periods are rational values. In actual
practice, the values of the grating periods are not integers but may be approximated by rational numbers. For
example, let $(\kappa_1,\kappa_2)=(194.7,13.3)$. In this particular case, we
can write $\kappa_1=p \kappa_2/133$, where $p$ is any integer which has no
common divisor with 133 except 1, showing that we are thus dealing with a
133'rd order resonance. In this particular case, $p=1947$, and $s$ must be
summed over $\pm133,\pm266,\pm399, \cdots$.

\begin{figure}
\centerline{\hbox{
\psfig{figure=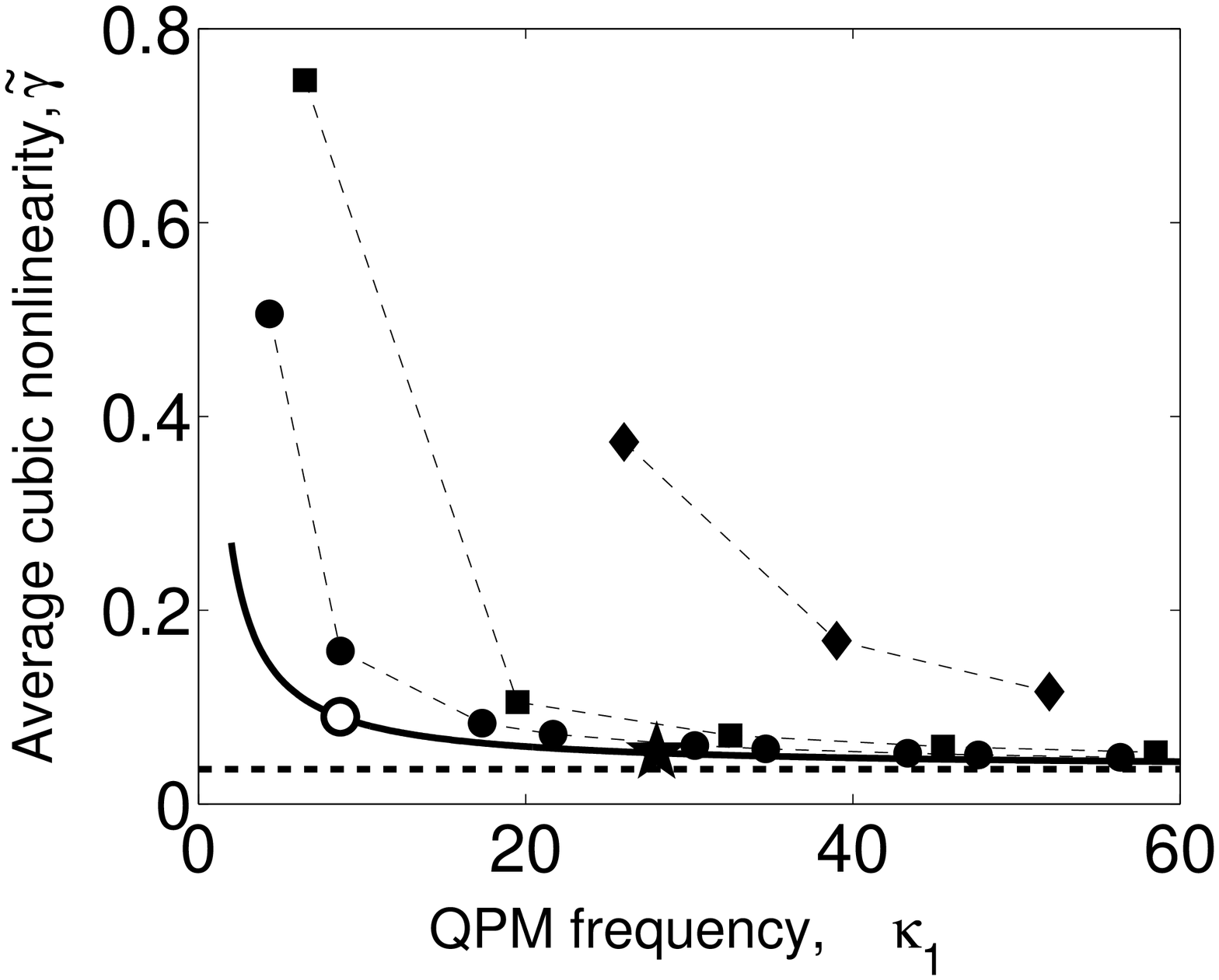,width=6cm,height=5cm}
}}\vspace{0.5cm}
\caption{Averaged cubic nonlinearity at the ($m=n=1$)-peak in the reduced system as a
function of the QPM frequency $\kappa_1$ ($\kappa_2=13$). Full
curve: No-resonance limit. $\scriptstyle{\blacklozenge}$: 1'st order resonances.
$\scriptscriptstyle{\blacksquare}$: 2'nd order resonances.
$\bullet$: 3'rd order resonances. The lines between
$\bullet$, $\scriptstyle{\blacklozenge}$ and
$\scriptscriptstyle{\blacksquare}$'s are only to help the eye. Dashed line:
$\kappa_2\rightarrow\infty$ approximation. $\circ$: Indicates a
point, $(\kappa_1,\tilde{\gamma})=(8.66,0.090)$ very close to the $p=2$
3'rd order resonance point
$(\kappa_1,\tilde{\gamma})=(26/3,0.157)$. $\bigstar$: Indicates the
point where $\tilde{\gamma}\approx0.050$ (see text).}
\label{fig:gamma13kap2}
\end{figure}

In Fig. \ref{fig:gamma13kap2} we show the induced averaged cubic nonlinearity
of the reduced system, $\tilde{\gamma}=\gamma/\eta^2$, as a function
of the QPM frequency $\kappa_1$ ($\kappa_2$ is fixed). We have plotted the values for
the lowest order resonances together with the no-resonance curve, i.e. the
limiting curve where the average nonlinearities are given by
(\ref{eqn:eta1}-\ref{eqn:gamma1}) which in the case of square grating
functions reduces to
\begin{eqnarray}
\eta   &=&-\frac{4}{\pi^2},\label{eqn:pracnl1}\\
\gamma&=&\frac{4}{\pi^2}\left(1-\frac{8}{\pi^2}\right)\left(\frac{1}{\kappa_1}+\frac{1}{\kappa_2}\right).\label{eqn:pracnl2}
\end{eqnarray}

It is evident from Fig. \ref{fig:gamma13kap2} that the higher order resonances
quickly converge towards the no-resonance curve with $\kappa_1$ and that
only the first order resonances yield induced averaged cubic nonlinearity
significantly higher than the no-resonance case. We note that the resonance
peaks shown are truly discrete. Furthermore, when the grating periods are
non-integer numbers, only very high-order resonances occur.  The point labelled
$\circ$ in Fig.~\ref{fig:gamma13kap2} illustrates the idea. It corresponds to
$(\kappa_1,\kappa_2)=(8.66,13)$. While the very similar structure corresponding
to $(\kappa_1,\kappa_2)=(26/3,13)$ yields a third-order resonance contribution,
$(\kappa_1,\kappa_2)=(8.66,13)$ yields a 650th order resonance. Hence, the
negligible deviation relative to the no-resonance curve.  The conclusion is
thus that only the no-resonance curve is of practical experimental interest.

\begin{figure}
\centerline{\hbox{
\psfig{figure=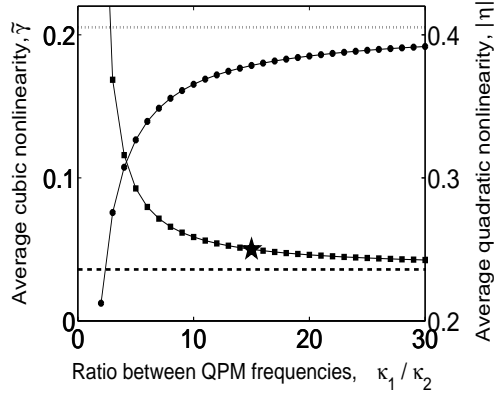,width=6cm,height=5cm}
}}\vspace{0.5cm}
\caption{Averaged nonlinearities at the first order resonances from Fig.
\ref{fig:gamma13kap2}. $\kappa_2=13$ and $\kappa_1=p*13$ ($p$ is an integer).
$\scriptscriptstyle{\blacksquare}$: The effective cubic nonlinearity,
$\tilde{\gamma}$, of the averaged system. $\bullet$: Amplitude of the averaged
quadratic nonlinearity, $\eta$. The curves are discrete and the lines in
between the $\bullet$'s and $\scriptscriptstyle{\blacksquare}$'s are just to
help the eye. The dashed and dotted lines indicate the asymptotic values of
$\tilde{\gamma}$ ($\kappa_1\rightarrow\infty$) and $|\eta|$, respectively.
$\bigstar$ indicates the point where $\tilde{\gamma}\approx0.050$ (see text).}
\label{fig:nonlin}
\end{figure}

\begin{figure}
\centerline{\hbox{
\psfig{figure=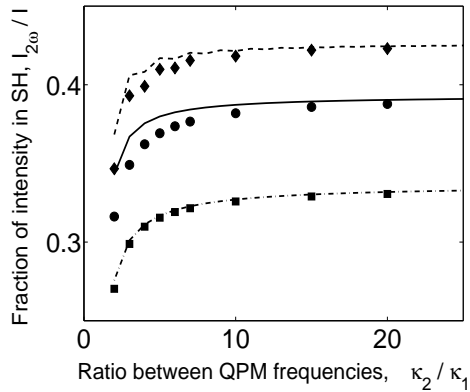,width=6cm,height=5cm}
}}\vspace{0.5cm}
\caption{Fraction of peak-intensity in SH at the first order resonances from
Fig. \ref{fig:nonlin}. $\kappa_2=13$ and $\kappa_1=p*13$ ($p$ is an
integer).  The input power is kept constant at $P=15$ and $\epsilon=0$.
The lines represent theoretically predicted values whereas
$\scriptstyle{\blacklozenge}$, $\bullet$ and
$\scriptscriptstyle{\blacksquare}$ are numerically measured values. 
Dashed line and $\scriptstyle{\blacklozenge}$: Maximum value. Solid
line and $\bullet$: Average value. Dashed dotted line and
$\scriptscriptstyle{\blacksquare}$: Minimum value. Notice that the
lines actually represent the discrete integer values. They are drawn
as lines just to help the eye.}
\label{fig:intkapfig}
\end{figure}

However, to verify that the model is correct we still need to analyze
the full resonant structure of the averaged nonlinearities. Thus 
in Fig. \ref{fig:nonlin} we
focus on the first order resonances and show how the effective averaged
quadratic nonlinearity changes with the ratio $\kappa_1/\kappa_2$. In the same figure we have also plotted $\tilde{\gamma}$
to emphasize the interplay between $\tilde{\gamma}$ and $\eta$. Recall that the
graphs are not continuous but discrete functions of $\kappa_1/\kappa_2$, i.e.  the plot only holds for values of $\kappa_1$
which are integers of $\kappa_2$ ($\kappa_2=13$ in this case).
In Fig. \ref{fig:intkapfig} we have made a series of numerical experiments to
confirm that the averaged model derived in this paper does predict correct
behavior also at the QPM frequency resonances. Again we launch soliton found from
(\ref{eqn:redavgQPMsystem1}-\ref{eqn:redavgQPMsystem2}) in the real
system described by
(\ref{eqn:chi23system1}-\ref{eqn:chi23system2}), but now we keep the input
power fixed, while we change $\kappa_1$ in integer steps of $\kappa_2$.
Simulations where carried out in the  actual two-period QPM structure.  As before
we let the evolution continue until any initial transient has died out.  Then
we sample the minimum, maximum, and averaged peak intensities and compare those
with the theoretically predicted values.

\begin{figure}
\centerline{\hbox{
\psfig{figure=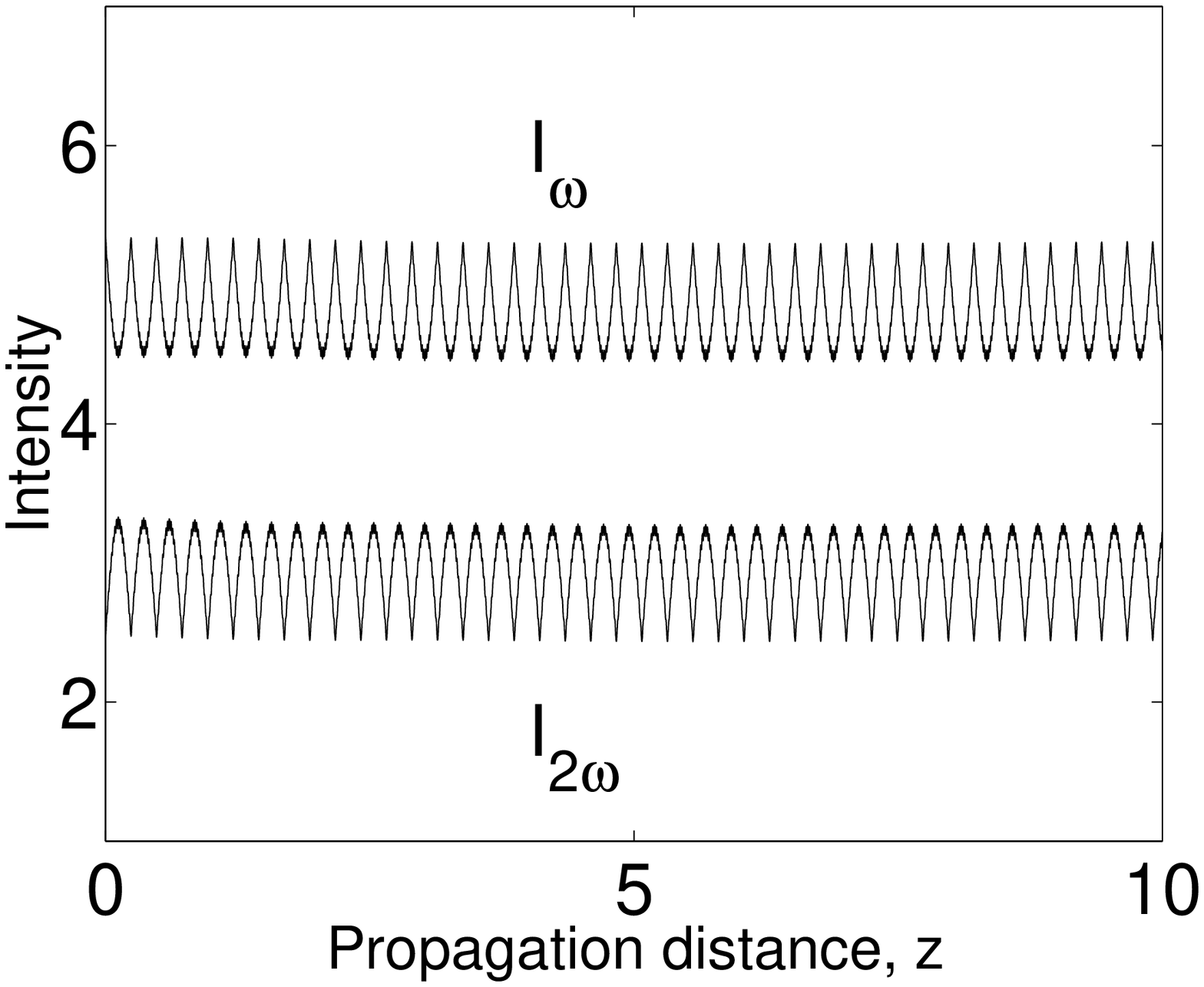,width=4cm,height=3.3cm}
\psfig{figure=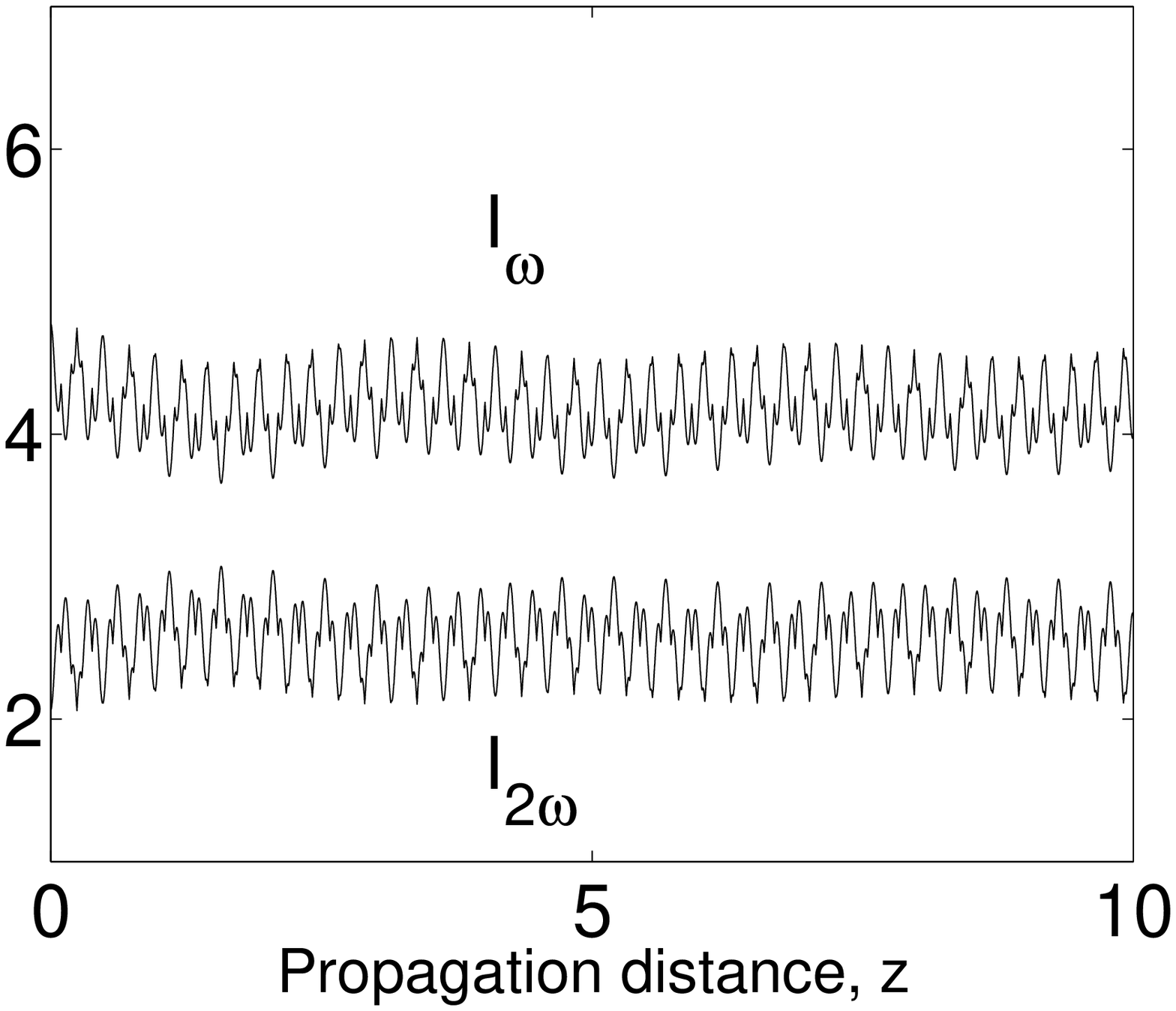,width=4cm,height=3.3cm}}}
\vspace{0.5cm}\centerline{\hbox{
\psfig{figure=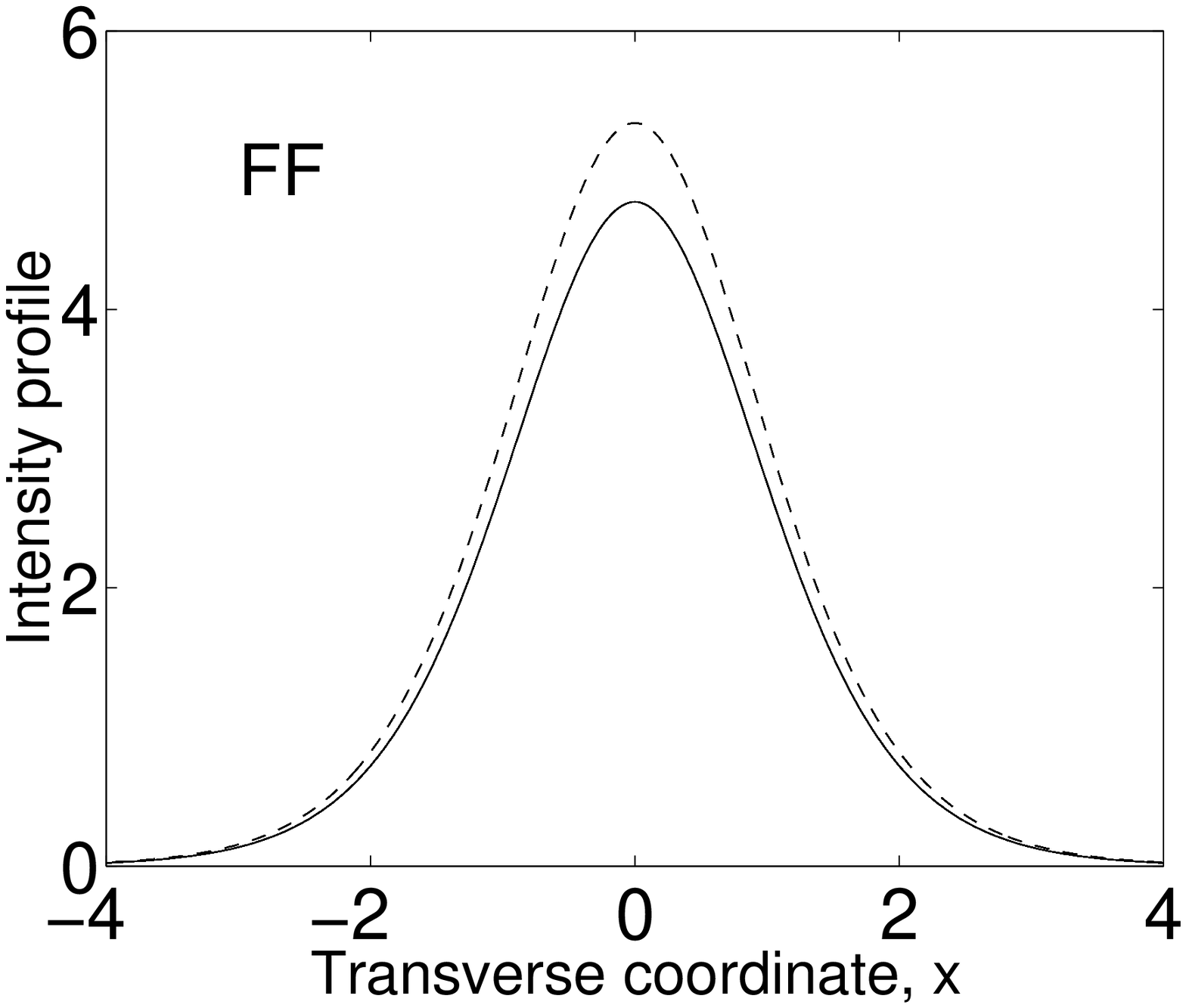,width=4cm,height=3.3cm}
\psfig{figure=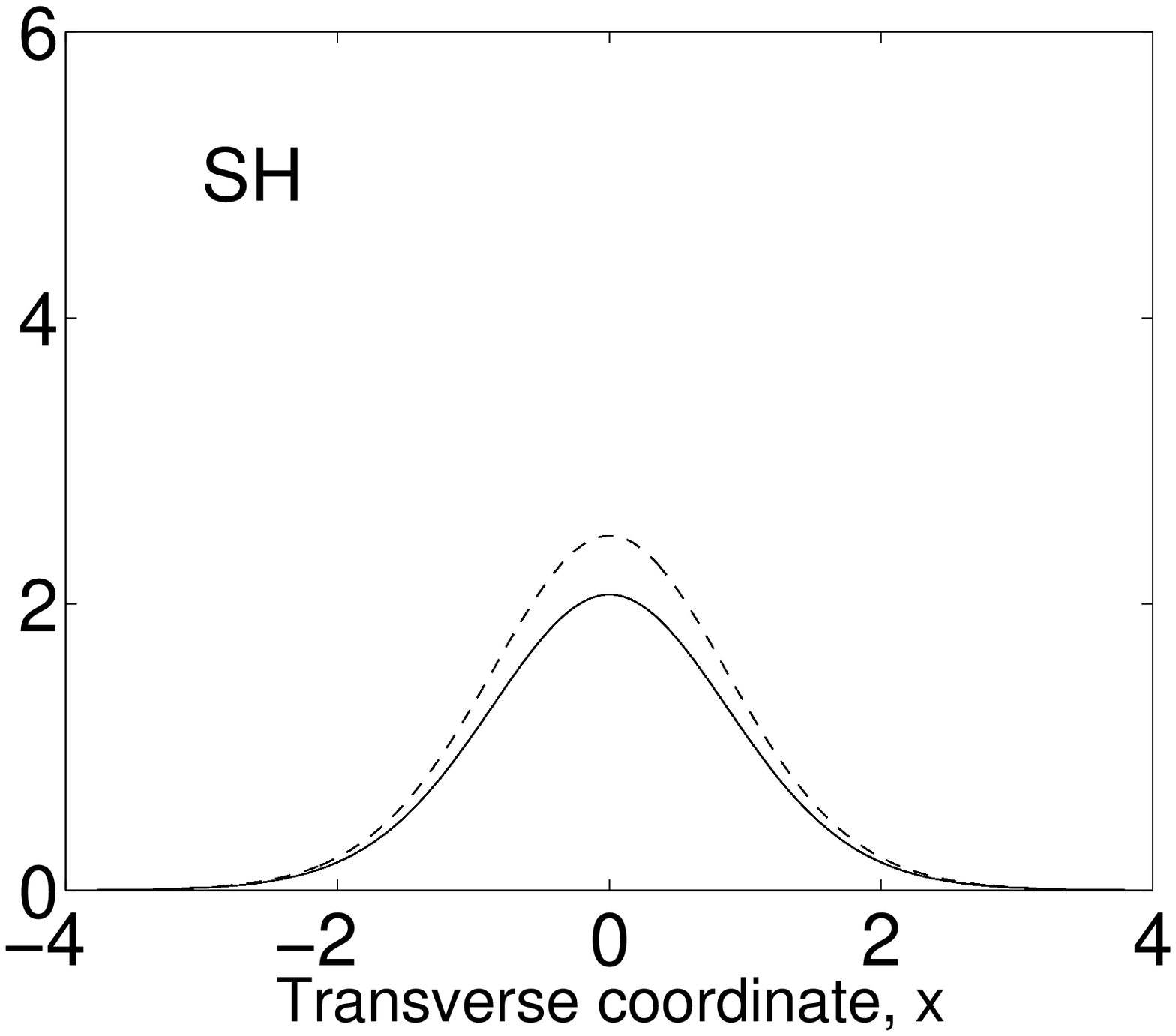,width=4cm,height=3.3cm}}}
\vspace{0.5cm}
\caption{Evolution of solitons in the real two-period system (\ref{eqn:chi23system1}-\ref{eqn:chi23system2}). Top
figures show peak intensities. Bottom figures compare intensity
profiles for the FF's (left) and SH's (right). Top left figure and dashed
curves in bottom figures are for $(\kappa_1,\kappa_2)=(195,13)$. Top
right figure and solid
curves in bottom figures are for $(\kappa_1,\kappa_2)=(32.9,13)$). The
two initial conditions stem from the same soliton of system
(\ref{eqn:redavgQPMsystem1}-\ref{eqn:redavgQPMsystem2}) with
$\tilde{\gamma}\approx0.050$, $\epsilon=0$, and $\lambda=0.4$.}
\label{fig:profil}
\end{figure}

It is worth recalling that for every soliton solution of the reduced system
(\ref{eqn:redavgQPMsystem1}-\ref{eqn:redavgQPMsystem2}) we have a whole family
of solutions of the form (\ref{eqn:newexpansion1}-\ref{eqn:newexpansion2}). In
Fig. \ref{fig:profil} we have shown how one specific solution for
$\tilde{\gamma}\approx0.050$ leads to different initial conditions for two
different physical setups. The two setups here chosen corresponds to the point
$\bigstar$ indicated on  both Fig. \ref{fig:gamma13kap2} and Fig.
\ref{fig:nonlin}. We also show the evolution of these initial conditions in
Fig. \ref{fig:profil}, i.e. the evolution of the peak intensities.

\begin{figure}
\centerline{\hbox{
\psfig{figure=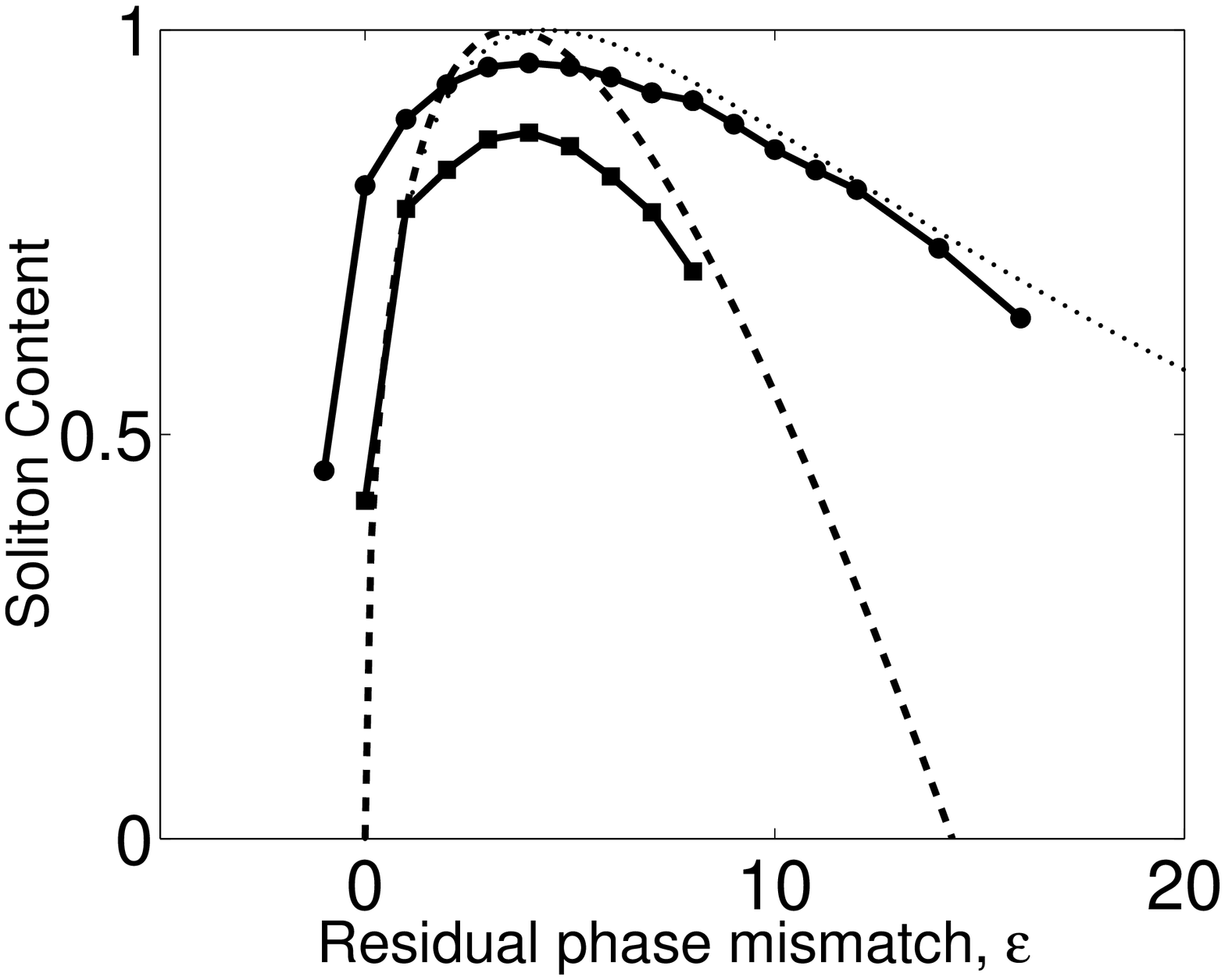,width=4cm,height=3.3cm}
\psfig{figure=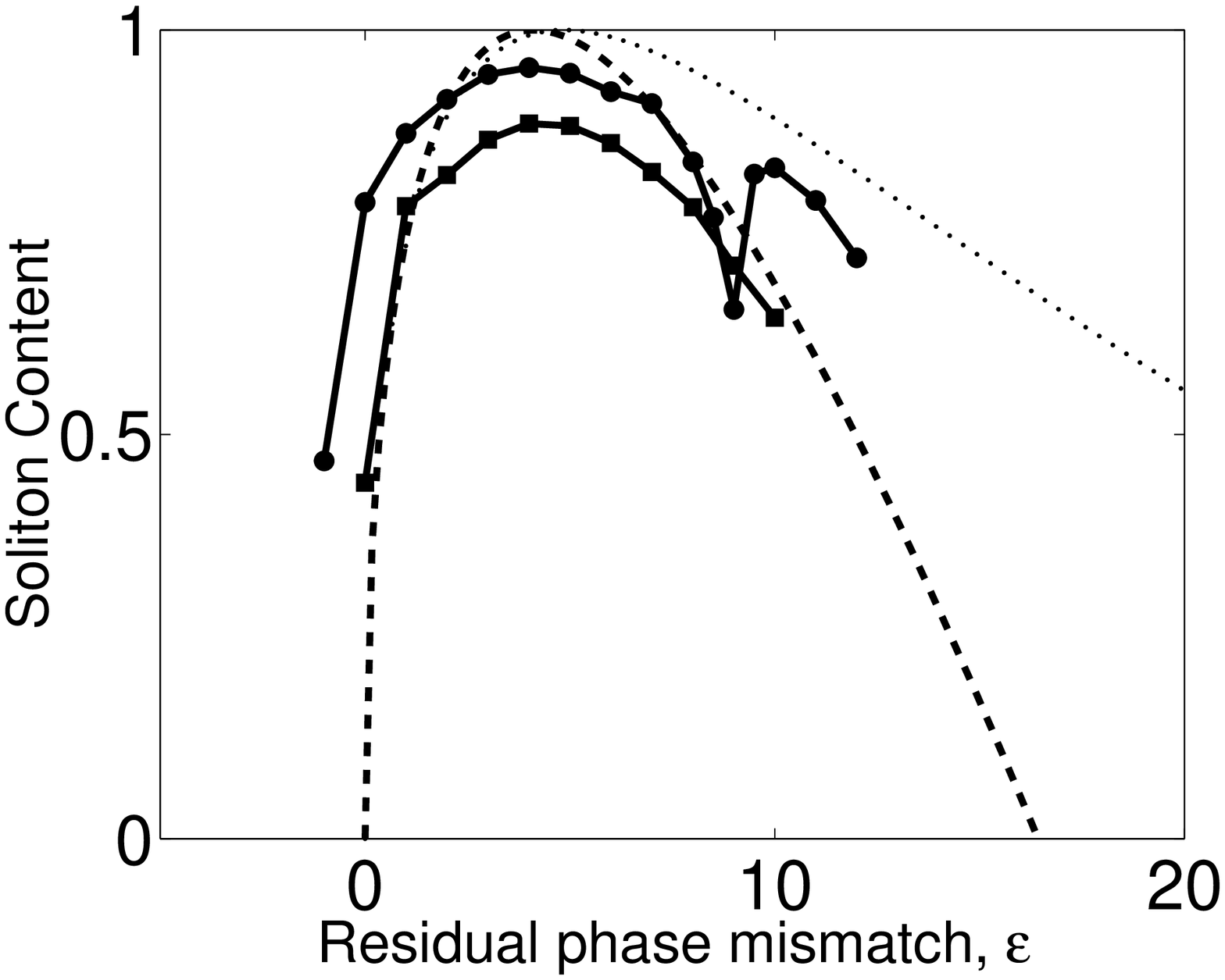,width=4cm,height=3.3cm}}}
\caption{Soliton content for sech input FF as a function of the residual
phase mismatch $\epsilon=\beta-\kappa_1-\kappa_2$ ($m=n=1$). Input power is
$P_{in}=50$. Left plot is for $(\kappa_1,\kappa_2)=(195,13)$
($\eta\simeq-0.38,\tilde{\gamma}\simeq0.050$) and right
plot for $(\kappa_1,\kappa_2)=(194.7,13.3)$  ($\eta\simeq-4/\pi^2,\tilde{\gamma}\simeq0.038$). $\bullet$': SC for two-period QPM. '$\scriptstyle{\blacksquare}$': SC for purely quadratic
model. The discrete points are the outcome of numerical experiments; the
full lines in between are only to help the eye. The dashed and dotted
lines are estimates from the limiting NLSE for a pure quadratic model
and for a model with cubic terms, respectively.}
\label{fig:solcont}
\end{figure}

For the induced averaged nonlinearities addressed in this paper to be of
potential practical importance, they have to impact the observable soliton
properties, including their excitation conditions. To show that this seems to
be the case, in Fig. \ref{fig:solcont} we show the behavior found for the
soliton content\cite{torner99oc}, SC, as a function of the residual phase
mismatch in a two-period structure with QPM frequencies
$(\kappa_1,\kappa_2)=(195,13)$. We also include the outcome of the numerical
experiments performed for a structure with the non-integer values
$(\kappa_1,\kappa_2)=(194.7,13.3)$, to show that the effect predicted is not
restricted to a particularly suitable choice of these parameters. We launch a
FF signal, with a sech-shape, and no SH seeding and calculate how much of the
initial power, $P_{in}$, is bound in the soliton which eventually
forms. We propagate until a steady state has emerged and then we collect the
energy flow in a window wide enough to enclose substantially all the
soliton. Typical values are propagation until $z=10^3$ and collection
of energy flow in a window of $x=\pm10$, but these values were adapted
whenever needed in order to capture always all the soliton energy. Simulations
where carried out in the actual two-period QPM structure. The bandwidth of the
SC for sech inputs can be estimated by using the Zakharov-Shabat scattering
equations associated with the (1+1)-dimensional nonlinear Schr\"odinger
equation, NLSE. With cubic terms, i.e. for system
(\ref{eqn:avgQPMsystem1}-\ref{eqn:avgQPMsystem2}), one gets the estimate
\begin{eqnarray}
\mbox{SC}\simeq
\frac{2 \sqrt{\frac{\eta^2\epsilon}{\eta^2+\gamma\epsilon}}}{\eta^2P_{in}}\left(\sqrt{2\eta^2P_{in}}-\sqrt{\frac{\eta^2\epsilon}{\eta^2+\gamma\epsilon}}\right).\label{eqn:scchi23}
\end{eqnarray}
In the plot we also include the actual value of SC that is numerically found in
the corresponding homogeneous pure quadratic case. Compared to the
pure quadratic model, the bandwidth for the two-period structure is
found to get wider. This effect can be understood by
comparing the approximate theoretical expressions for the two cases
given by (\ref{eqn:scchi23}). There is also an expected significant increase in
the efficiency owing to the fact that in systems with induced cubic
terms the exited solitons are closer to the initial sech-shape than in
purely quadratic systems. To explain the ``dip'' in the SC in the case of
$(\kappa_1,\kappa_2)=(194.7,13.3)$ again one must turn to the SHG
spectrum. We find that the $(m=-1,n=31)$-peak is located exactly at
$\epsilon=9.6$, which is where the bottom of the ``dip'' seems to
be. Thus we conclude that the efficiency of soliton excitation is
diminished because of resonances between the $(m=n=1)$- and the
$(m=-1,n=31)$-peak.

\begin{figure}
\centerline{\hbox{
\psfig{figure=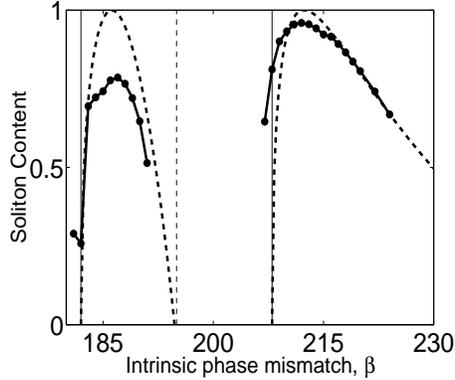,width=6cm,height=5cm}}}
\caption{Soliton content for sech input FF as a function of intrinsic
phase mismatch $\beta$ with $(\kappa_1,\kappa_2)=(195,13)$. Input power is
$P_{in}=50$. The discrete points are the outcome of numerical experiments; the
full lines in between are only to help the eye. The dashed curves are
estimates from the limiting NLSE. The vertical lines located at
$\beta=182$ and $\beta=208$ indicate the $m=1,n=-1$ and $m=n=1$ peaks,
respectively. The dashed vertical line at $\beta=195$ indicates the
location of the peak in the absence of a second period (see Fig. \ref{fig:peaks}).}
\label{fig:k19513sc}
\end{figure}

\begin{figure}
\centerline{\hbox{
\psfig{figure=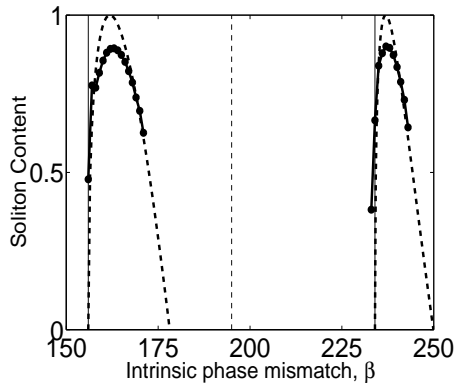,width=6cm,height=5cm}}}
\caption{Same as in Fig. \ref{fig:k19513sc}, but here
 $(\kappa_1,\kappa_2)=(195,39)$}. 
\label{fig:k19539sc}
\end{figure}

In Fig. \ref{fig:k19513sc} we plot the soliton content for the
$(\kappa_1,\kappa_2)=(195,13)$ case, but now we scan mismatches not
only around the ($m=n=1$)-peak but also around the $(m=1,n=-1)$-peak. In the
absence of resonances and average cubic nonlinearities, i.e. $\eta=
\pm 4/\pi^2$, no differences are found in the behavior of soliton
excitation at each peak. However, such is not the case when the soliton
generation efficiencies can interfere with each other, as shown in Fig.
\ref{fig:k19513sc}. For example, in the particular case shown, one observes
that soliton generation around the $(m=1,n=-1)$-peak takes place within a
narrower band of mismatches and is less efficient than around the
($m=1,n=1$)-peak. This is because the average nonlinearities
(\ref{eqn:squarenl1}-\ref{eqn:squarenl3}) are nonlinear functions of $m$ and
$n$ and hence they change their relative strengths at the two peaks. A
different behavior than that displayed in Fig. \ref{fig:k19513sc} might be
obtained with different values of the two-grating periods and input light
conditions, as shown in Fig. \ref{fig:k19539sc}. When resonances are involved we note that at the  $(m=1,n=-1)$-peak
the induced part of the average second order nonlinearity is stronger than the
no-resonance value $4/\pi^2$ whereas it is weaker at the ($m=1,n=1$)-peak.
Notice also that the soliton content vanishes in the intermediate region
between the ($m=1, n=\pm 1$)-peaks.  This is because in the intermediate region
between both peaks, e.g., at the mismatch that would be exactly compensated
with a single-period QPM structure, the interference between the two peaks acts
as a periodic amplitude modulation of the effective nonlinear coefficient, with
a period comparable with the characteristic soliton length.  Such modulation
constitutes a very strong perturbation that prevents soliton formation.

We remark that one finds soliton generation around other peaks than the
($m=1,n=\pm1$)-peak. For example, in the $(\kappa_1,\kappa_2)=(195,13)$ case we
observe a band (not shown in the plot) around $\epsilon=39$ corresponding to
the ($m=1,n=3$)-peak. However, soliton formation in this band is much less
efficient than in the ($m=1,n=\pm1$)-band and therefore we did not include it
the plot.  Similarly, other bands with higher QPM order also exist but one has
to launch correspondingly high powers to excite solitons.

In conclusion, we derived an averaged model for the soliton propagation in
two-period QPM systems of the form
(\ref{eqn:avgQPMsystem1}-\ref{eqn:avgQPMsystem2}), and we showed that the model
gives an accurate description of stationary soliton propagation under a variety
of input light and material conditions. We have shown that one can use one
grating period of the QPM structure to reduce the intrinsic phase mismatch and
the other grating period to induce averaged cubic nonlinearities. We found the
induced averaged cubic nonlinearities to be strongest when one QPM frequency is a
multiple of the other, but we showed that those are strictly discrete
resonances that cannot occur when experimentally feasible grating periods are
taking into account. However, our investigations predict that also the no-resonance
averaged contributions can be important. In particular, we found that the
QPM engineered averaged cubic nonlinearities, induced in feasible two-period
samples, enhances the peak-efficiency and residual mismatch-bandwidth of the
soliton excitation process with non-soliton single frequency pump light. 

\noindent
\textbf{Acknowledgements}\\
This work was supported by the European Union through the Improving Human
Potential program (contract HPRI-CT-1999-00071). Silvia Carrasco and Lluis
Torner acknowledge support from the Generalitat de Catalunya, and by the
Spanish government under TIC2000-1010. Ole Bang acknowledges support from the
Danish Technical Research Council under Talent Grant No. 26-00-0355. Numerical
work was carried out at CESCA-CEPBA-CIRI.
\vspace{.5cm}

\noindent
$^{*}$ Permanent address: Department of Informatics and Mathematical
Modelling, Technical University of Denmark, DK-2800 Lyngby, Denmark.

\end{document}